\documentclass[sigconf,nonacm]{acmart}

\usepackage{graphicx}
\usepackage{booktabs}
\usepackage{arydshln}
\usepackage{fontawesome5}
\usepackage{balance}
\usepackage{multirow}
\usepackage[inline, shortlabels]{enumitem}

\AtBeginDocument{%
  \providecommand\BibTeX{{%
    \normalfont B\kern-0.5em{\scshape i\kern-0.25em b}\kern-0.8em\TeX}}}

\usepackage{collcell}
\usepackage{booktabs}

\usepackage{color, colortbl}
	
\definecolor{Gray}{gray}{0.9}

\def\highlightifnumstyle{\bfseries}
\newcommand\highlightifnum[3]{%
  \ifnum#1#2#3%
    {\highlightifnumstyle #3}%
  \else%
    #3%
  \fi%
}

\newcolumntype{H}[3]{%
  >{%
    \def\wrapper{\highlightifnum{#1}{#2}}%
    \collectcell\wrapper%
  }%
  #3%
  <{\endcollectcell}%
}

\copyrightyear{2023}
\acmYear{2023}
\setcopyright{acmlicensed}\acmConference[SIGIR '23]{Proceedings of the 46th International ACM SIGIR Conference on Research and Development in Information Retrieval}{July 23--27, 2023}{Taipei, Taiwan}
\acmBooktitle{Proceedings of the 46th International ACM SIGIR Conference on Research and Development in Information Retrieval (SIGIR '23), July 23--27, 2023, Taipei, Taiwan}
\acmPrice{15.00}
\acmDOI{10.1145/3539618.3591956}
\acmISBN{978-1-4503-9408-6/23/07}

\begin{document}

\title{A Thorough Comparison of Cross-Encoders and LLMs for Reranking SPLADE}

\author{Hervé Déjean, Stéphane Clinchant, Thibault Formal}
\email{ first.lastname@naverlabs.com}
\affiliation{%
  \institution{Naver Labs Europe}
  \streetaddress{6 chemin de Maupertuis}
  \city{Meylan}
    \country{France}
}

\renewcommand{\shortauthors}{Déjean et al.}
\begin{abstract}
We present a comparative study between cross-encoder and LLMs rerankers in the context of re-ranking effective SPLADE retrievers. We conduct a large evaluation on TREC Deep Learning datasets and out-of-domain datasets such as BEIR and LoTTE. 
In the first set of experiments, we show how cross-encoder rerankers are hard to distinguish when it comes to re-rerank SPLADE on MS MARCO. Observations shift in the out-of-domain scenario, where both the type of model and the number of documents to re-rank have an impact on effectiveness. 
Then, we focus on listwise rerankers based on Large Language Models -- especially GPT-4. While GPT-4 demonstrates impressive (zero-shot) performance, we show that traditional cross-encoders remain very competitive. Overall, our findings aim to to provide a more nuanced perspective on the recent excitement surrounding LLM-based re-rankers -- by positioning them as another factor to consider in balancing effectiveness and efficiency in search systems.

\end{abstract}

\keywords{Information Retrieval, Neural Search, Reranking, Cross-Encoders, Large Language Models}



\maketitle

\section{Introduction}

Reranking models significantly enhance the quality of Information Retrieval (IR) systems. Due to their complexity, they are usually bound to reorder a limited number of documents provided by an efficient first-stage retriever such as BM25~\cite{conf/trec/RobertsonWJHG94,10.1561/1500000019}. Traditional reranking methods used to rely on manually defined features, and employed specific learning-to-rank losses \cite{learning2rank_book}. Since the advent of models like BERT~\cite{Devlin2019BERTPO}, cross-encoders have become the standard reranking ``machinery'' \cite{nogueira2020passage,gao_rethink_2021}. More recent architectures have gradually been tested, including encoder-decoder \cite{pradeep2021expandomonoduo,zhuang_setwise_2023,10.1145/3539618.3592047} or decoder-only models \cite{ma_fine-tuning_2023}.   
More recently, Large Language Models (LLMs) have been shown to be effective \emph{zero-shot} rerankers. For instance, RankGPT~\cite{sun_is_2023} -- relying on OpenAI GPT-4 \cite{openai2024gpt4} -- provides puzzling outcomes: it performs very well as a listwise reranker out-of-the-box, and can even be iteratively re-applied to incrementally improve the reranked lists.


However, we notice that strong baselines are often absent or not systematically used in recent works evaluating LLM-based rerankers (e.g.,  \cite{sun_is_2023}). For instance, it remains unclear whether such approaches significantly outperform standard cross-encoders when re-ordering the results of strong retrievers -- and if so, in which setting (e.g., how many documents to consider). Therefore, this study aims to shed light on such questions, and in particular:
\begin{itemize}
    \item Provide a comprehensive evaluation (in both in-domain and out-of-domain settings) of cross-encoder-based and LLM-based rerankers, when it comes to re-ordering effective retrievers (namely, the latest series of SPLADE-v3 models~\cite{splade2023}). 
    \item Evaluate the impact of various re-ranking settings on effectiveness. 
    \item Thoroughly compare RankGPT with a state-of-the-art cross-encoder.
    \item Evaluate open LLMs as (zero-shot) listwise rerankers.
\end{itemize}


Therefore, we conduct an extensive experimental study on the TREC Deep Learning datasets (19-23)~\cite{craswell2020overview,Craswell2022OverviewOT,Craswell2021OverviewOT,craswell2024overview} for in-domain evaluation, as well as the BEIR~\cite{Thakur2021BEIRAH} and LoTTE~\cite{Santhanam2021ColBERTv2EA} collections for out-of-domain evaluation.
Overall, it is difficult to draw general conclusions from this extensive evaluation, but our findings reveal that:
\begin{itemize}
    \item Cross-encoder rerankers behave slightly differently on in-domain and out-of-domain datasets.
    \item Cross-encoders remain competitive against LLM-based re-rankers -- in addition to being way more efficient.
    \item Open LLMs such under-perform compared to GPT-4, but still exhibit good ranking abilities under some constraints (e.g., small prompts).
\end{itemize}

\section{LLMs as Rerankers}

RankGPT \cite{sun_is_2023} is the first approach to investigate the direct use of LLMs as rerankers -- the model \emph{generating} an ordered list of document {\it ids} as output. To bypass the inherent prompt length limit of GPT models, {\it Sun et al.} introduce a sliding window strategy that allows LLMs to rank an arbitrary number of passages. Both GPT-3.5 Turbo and GPT-4 \cite{openai2024gpt4} are evaluated -- the latter providing remarkable results (especially given the zero-shot nature of the task). Later on, {\it Tang et al.}~\cite{tang_found_2023} present a more effective approach (PSC) to rerank the documents by comparing permuted input lists.

Experiments using open-source LLMs -- Open AI LLMs being closed and sometimes very expensive -- are however more underwhelming: {\it Qin et al.}~\cite{qin_large_2023} show \emph{that listwise ranking approaches generate completely useless outputs on [open] moderate-sized LLMs}. They therefore propose a pairwise approach with an advanced sorting algorithm (PRP-Sorting) to improve and speed up reranking. {\it Zhuang et al.}~\cite{zhuang_setwise_2023} further propose to compare the different ways to conduct reranking: pointwise, pairwise, and listwise. They introduce a new setwise prompting method which leads to a more effective and efficient listwise (zero-shot) ranking. Their experiments are conducted by fine-tuning a FLAN-T5-XXL model~\cite{chung2022scaling}. Besides, this work seems to be the first one really questioning the (un)efficiency of LLMs as rerankers. 

In the meantime, several approaches have studied \emph{fine-tuning} of LLMs for the task of reranking~\cite{zhang_rankinggpt_2023, ma_fine-tuning_2023, pradeep_rankzephyr_2023}.
For instance, {\it Pradeep et al.}~\cite{pradeep_rankzephyr_2023} fine-tune a moderate-sized LLM based on the Zephyr-7$B$ model~\cite{tunstall2023zephyr}, and achieve competitive results on par with GPT-4. 

While demonstrating impressive capabilities (both in zero-shot or fine-tuning settings), these models are relatively inefficient\footnote{Reranking 50 documents can take up to 1 minute using GPT-4 and/or Llama-70$B$ on an $H100$ GPU.} compared to traditional rerankers based on cross-encoders -- which are themselves considered as a ``slow'' components in IR systems. How to make these LLMs more efficient remains unclear.

\section{Experimental Setting}

We describe our experimental setting: the retrievers we consider, and the cross-encoder and LLM-based re-rankers we aim to compare in in- and out-of-domain settings.  

\subsection{An Effective Retriever -- SPLADE-v3}
\label{sec:splade}

While demonstrating impressive results, rerankers based on LLMs haven't been thoroughly evaluated when coupled with more effective first-stage models\footnote{Except some works like \cite{pradeep_rankzephyr_2023,ma_fine-tuning_2023}.}. For instance, RankGPT \cite{sun_is_2023} solely re-orders BM25 top documents. However, state-of-the-art retrievers such as SPLADE++ \cite{formal2022distillation} or SPLADE-v3~\cite{splade2023} already achieve \textit{better} results than the ones presented in RankGPT. Obviously, a good first-stage retriever ``makes things easier'' for the reranker, but it is still unclear how they interact. 

We, therefore, propose to study reranking for highly effective retrievers -- and focus on SPLADE models due to their good results on several tracks of the TREC Deep Learning evaluation campaign \cite{lassance2023naver} as well as out-of-domain benchmarks. More specifically, we test the three variants of the latest series of SPLADE models\footnote{The SPLADE-v3 checkpoints are available on HuggingFace -- see Table\ref{tab:models}}~\cite{splade2023}: 
\begin{enumerate*}
    \item SPLADE-v3, 
    \item SPLADE-v3-DistilBERT,
    \item SPLADE-v3-Doc. 
\end{enumerate*}


While SPLADE-v3 outperforms its two more efficient counterparts, we aim to assess the impact of reranking on the final performance with ``weaker'' models.

\subsection{Rerankers Based on Cross-Encoders}
\label{sec:rerankers}

We then evaluate two cross-encoders specifically trained to re-rank SPLADE models~\cite{lassance2023naver,Gao2021RethinkTO}. Specifically, we select the ones based on DeBERTa-v3 large \cite{he2023debertav3} and ELECTRA-large \cite{clark2020electra}\footnote{The rerankers are available on HuggingFace -- see Table\ref{tab:models}}.
The DeBERTa-v3 large model comes with 24 layers and a hidden size of 1024. It has 304$M$ parameters, and a Byte-Pair-Encoding vocabulary containing 128$k$ tokens.
The ELECTRA model has similar specifics -- besides the WordPiece vocabulary containing $30k$ tokens -- with 335$M$ parameters in total.

\subsection{Under the Hood of RankGPT}

Following {\it Sun et al.}~\cite{sun_is_2023}, we use GPT-3.5 and GPT-4 as \emph{zero-shot} rerankers. To bypass the limited prompt length of GPT, we also use the sliding window strategy that allows ranking a larger pool by iteratively ranking overlapping chunks of documents. We show (Section~\ref{sec:4-3}) that using a more effective retriever (in our case, SPLADE-v3) allows us to rerank a smaller number of documents without resorting to this window strategy.

Note that Open AI models -- especially GPT-4 -- are rather expensive. Due to budget limits, some of our experiments are not as comprehensive as we would have liked them to be. 

For the experiments, we rely on the RankGPT code\footnote{\url{https://github.com/sunnweiwei/RankGPT}}, and modify it to be able to use other (open) models. We now describe in detail the pre- and post-processing steps used in RankGPT -- which are critical aspects for such re-ranking approaches based on generative models.

\subsubsection{Sliding Window Strategy}

A key concern when using LLM-based rerankers is the number of documents their prompt can ingest. A key mechanism used by RankGPT is a sliding window strategy: the system starts by ranking the $N$ last documents, then creates an $k$-length overlapping window with the $N$ previous documents, ranks them, and iterates the process until reaching the $N$ first documents. Because ranking has to be performed several times per query, this approach is rather costly and inefficient. We show in the experiments that, depending on the model and the dataset, this mechanism is either useful or can be ignored.

\subsubsection{Pre-Processing: Truncating Documents}
\label{sec:shorten}

Another mechanism used to manage the prompt size is document truncation. As long as the $N$ documents do not fit into the prompt, documents are shortened by one token. The default document length in RankGPT is $|d|=300$ (which, however, is already quite large when considering MS MARCO passages). It is much more efficient than the window mechanism, but can negatively impact effectiveness.

\subsubsection{Post-Processing}

The output of the model -- assuming the instructions are followed by the LLM -- is a list of document identifiers ordered by decreasing relevance, and separated by a ``\texttt{>}'' token. For strong LLMs such as GPT-4, almost no post-processing is needed, but it is strongly recommended to use it with other models that do not always respect the formatting instructions. Additionally, document identifiers that are not present in the LLM output are added to the final output (according to the original ordering). It allows an LLM that does not generate anything ``meaningful'' to perform as well as the retriever. 

\subsubsection{Prompting}

{\it Pradeep et al.} \cite{pradeep_rankzephyr_2023} use the prompt designed for RankGPT -- we also follow the same strategy for all the LLMs in this study.

\subsection{Datasets}

For evaluation, we use two types of datasets: in-domain and out-of-domain. For in-domain datasets we use all the available TREC Deep Learning datasets based on the MS MARCO collection (passages): from DL19 to DL23 \cite{craswell2020overview,Craswell2022OverviewOT,Craswell2021OverviewOT,craswell2024overview}. 
For the cross-encoder rerankers, we use BEIR \cite{Thakur2021BEIRAH} -- with the 13 readily available datasets -- and LoTTE \cite{Santhanam2021ColBERTv2EA} as out-of-domain datasets. For the OpenAI LLMs rerankers, their cost prevents us from evaluating them at this scale. Therefore, we select datasets from BEIR for which the number of queries is around 50: TREC-COVID, TREC-NEWS, and Touché-2020. Finally, we also consider the NovelEval dataset~\cite{sun_is_2023} to assess GPT-4's effectiveness on unseen data (data issued after GPT-4 training).

\section{Results}

We first discuss the results with cross-encoders for in-domain and out-of-domain evaluation settings. Then we compare cross-encoders with LLMs rerankers. Finally, we conduct various experiments on the TREC-COVID dataset, to highlight the different behavior between cross-encoders and LLMs.  

Our results show that it isn't obvious to draw clear conclusions, as observations are quite dependent on datasets. 

In the following, we report nDCG@10\footnote{Multiplied by 100 to improve readability}. Note that we haven't (yet) performed \textit{statistical significance tests} -- but given the small number of queries in the DL datasets, a difference of less than 1 point nDCG@10 is usually not considered significant.
\begin{table*}[ht]
\centering
\caption{
Effectiveness of three SPLADE-v3 models on TREC DL datasets.  SPLADE-v3-DistilBERT serves as the reference point (\colorbox{Gray}{grey area}, nDCG@10) -- and the comparisons are given in $\Delta$(nDCG@10). The baselines correspond to the retrievers only (no re-ranking). 
}
\resizebox{0.9\textwidth}{!}{
\begin{tabular}{ll >{\columncolor{Gray}}c>{\columncolor{Gray}}c>{\columncolor{Gray}}c>{\columncolor{Gray}}c>{\columncolor{Gray}}cccccc ccccc ccccc}
\toprule
\multicolumn{2}{l}{} &  \multicolumn{5}{c}{\textbf{SPLADE-v3-DistilBERT}}   & \multicolumn{5}{c}{\textbf{SPLADE-v3}} &  \multicolumn{5}{c}{\textbf{SPLADE-v3-Doc}}   \\ 
\cmidrule(lr){3-7}  \cmidrule(lr){8-12} \cmidrule(lr){13-17} 
\multicolumn{2}{l}{} 
&  \cellcolor{white}\footnotesize{ \texttt{DL19}}
&  \cellcolor{white} \footnotesize{ \texttt{DL20}}
&  \cellcolor{white} \footnotesize{ \texttt{DL21}}
&  \cellcolor{white} \footnotesize {\texttt{DL22}}
&  \cellcolor{white} \footnotesize {\texttt{DL23}} 
&  \footnotesize {\texttt{DL19}}
&  \footnotesize{ \texttt{DL20}}
&  \footnotesize{ \texttt{DL21}}
&  \footnotesize {\texttt{DL22}}
&  \footnotesize {\texttt{DL23}} 
& \footnotesize{ \texttt{DL19}}
&  \footnotesize{ \texttt{DL20}}
&  \footnotesize{ \texttt{DL21}}
&  \footnotesize {\texttt{DL22}}
&  \footnotesize {\texttt{DL23}} \\
\midrule
\texttt{baseline} &  & 72.3 & 75.4&70.7& 61.9&50.7 &  -3.2 & 0.9 & 5.5 & 6.8 & 4.1 & -3.9 & -4.1 & -0.3 & -3.5 & -0.2 \\
\midrule
\textbf{Reranker} & {\textbf{top$_k$}} \\
\texttt{DeBERTa-v3}    & 50  &  78.1 & 75.3& 74.1 & 65.4 & 52.2 & -0.9 & 0.6 & 0.6 & 2.7 & 6.1 & -1.3 & 0.7 & -1.2 & -0.3 & 4.1 \\
&100 & 78.5 & 75.5 & 73.4 & 66.3 & 55.8 & -1.1 & 0.0 & 0.6 & 1.7 & 2.3 & -0.8 & -0.1 & -0.1 & 0.4 & 0.6 \\
&200 & 78.2 & 75.5 & 74.0 & 67.0 & 57.0 & -0.7 & 0.1 & 0.0 & 0.8 & 1.3 & -0.3 & 0.0 & -0.6 & 0.5 & -0.2 \\

\texttt{ELECTRA} & 50 &  77.5 & 77.1 & 74.5 & 64.1 & 55.5 & -0.6 & 0.6 & 0.2 & 3.9 & 2.4 & -0.8 & 0.1 & -1.3 & 0.4 & 1.4 \\
 & 100& 77.4 & 77.3 & 74.2 & 65.6 & 57.0 & -0.4 & 0.1 & -0.1 & 1.6 & 0.6 & -0.4 & 0.0 & -0.2 & 0.44 & 0.8 \\
 & 200& 76.8 & 77.5 & 74.0 & 66.6 & 57.2 & 0.4 & 0.0 & -0.1 & 0.0 & 0.4 & 0.2 & 0.0 & -0.1 & -0.4 & -0.3 \\

\bottomrule
\end{tabular}
}
\label{tab:table_1compmodel}
\end{table*}

\begin{table*}[ht]
\centering
\caption{
Comparison of DeBERTa-v3 and ELECTRA rerankers for the three SPLADE-v3 models. DeBERTa-v3 serves as the reference point (\colorbox{Gray}{grey area}, nDCG@10) -- and the comparisons are given in $\Delta$(nDCG@10).}
\resizebox{0.9\textwidth}{!}{
\begin{tabular}{ll cccccccccc ccccc ccccc}
\toprule
\multirow{2}{*}{\textbf{Reranker}}& \multirow{2}{*}{\textbf{top$_k$}} &\multicolumn{5}{c}{\textbf{SPLADE-v3-DistilBERT}}   & \multicolumn{5}{c}{\textbf{SPLADE-v3}} &  \multicolumn{5}{c}{\textbf{SPLADE-v3-Doc}}   \\ 
\cmidrule(lr){3-7}  \cmidrule(lr){8-12} \cmidrule(lr){13-17} 
\textbf{}
& \textbf{}
& \footnotesize{ \texttt{DL19}}
&  \footnotesize{ \texttt{DL20}}
&  \footnotesize{ \texttt{DL21}}
&  \footnotesize {\texttt{DL22}}
&  \footnotesize {\texttt{DL23}} 
&  \footnotesize{\texttt{DL19}}
&  \footnotesize{ \texttt{DL20}}
&  \footnotesize{ \texttt{DL21}}
&  \footnotesize {\texttt{DL22}}
&  \footnotesize {\texttt{DL23}} 
& \footnotesize{ \texttt{DL19}}
&  \footnotesize{ \texttt{DL20}}
&  \footnotesize{ \texttt{DL21}}
&  \footnotesize {\texttt{DL22}}
&  \footnotesize {\texttt{DL23}} \\
\midrule
\rowcolor{Gray}  \cellcolor{white}\texttt{DeBERTa-v3}    & \cellcolor{white} 50  &  78.1 & 75.3 & 74.1 & 65.4 & 52.2 & 77.2 & 75.9 & 74.7 & 68.2 & 58.3 & 76.8 & 76.0 & 72.9 & 65.1 & 56.4 \\
\rowcolor{Gray}  \cellcolor{white} & \cellcolor{white}100 & 78.5 & 75.5 & 73.4 & 66.3 & 55.8 & 77.4 & 75.5 & 74.0 & 68.0 & 58.2 & 77.7 & 75.4 & 73.2 & 66.8 & 56.3 \\
\rowcolor{Gray}  \cellcolor{white} & \cellcolor{white}200 & 78.2 & 75.5 & 74.0 & 67.0 & 57.0 & 77.5 & 75.6 & 74.0 & 67.9 & 58.3 & 77.8 & 75.5 & 73.4 & 67.4 & 56.9 \\

\texttt{ELECTRA} & 50 & -0.6 & 1.9 & 0.5 & -1.3 & 3.3 & -0.3 & 1.9 & 0.0 & -0.2 & -0.4 & -0.2 & 1.3 & 0.4 & -0.7 & 0.6 \\
&100 & -1.0 & 1.8 & 0.8 & -0.7 & 1.2 & -0.4 & 1.9 & 0.1 & -0.8 & -0.5 & -0.7 & 2.0 & 0.8& -0.7 & 1.4 \\
&200 & -1.4 & 2.0 & 0.0 & -0.4 & 0.2 & -0.3 & 1.9 & 0.0 & -1.2 & -0.7 & -0.8 & 1.9 & 0.5 & -1.3 & 0.1 \\

\bottomrule
\end{tabular}
}
\label{tab:table_1compreranker}
\end{table*}

\begin{table*}[ht]
\centering
\caption{
Impact of the number of documents to re-rank (top$_k$) on effectiveness. top$_k=50$ serves as the reference point (\colorbox{Gray}{grey area}, nDCG@10) -- and the comparisons are given in $\Delta$(nDCG@10).
}
\resizebox{0.95\textwidth}{!}{
\begin{tabular}{ll cccccccccc ccccc ccccc}
\toprule
\multirow{2}{*}{\textbf{Reranker}}& \multirow{2}{*}{\textbf{top$_k$}} &\multicolumn{5}{c}{\textbf{SPLADE-v3-DistilBERT}}   & \multicolumn{5}{c}{\textbf{SPLADE-v3}} &  \multicolumn{5}{c}{\textbf{SPLADE-v3-Doc}}   \\ 
\cmidrule(lr){3-7}  \cmidrule(lr){8-12} \cmidrule(lr){13-17} 
\textbf{}
& \textbf{}
& \footnotesize{ \texttt{DL19}}
&  \footnotesize{ \texttt{DL20}}
&  \footnotesize{ \texttt{DL21}}
&  \footnotesize {\texttt{DL22}}
&  \footnotesize {\texttt{DL23}} 
&  \footnotesize{DL19}
&  \footnotesize{ \texttt{DL20}}
&  \footnotesize{ \texttt{DL21}}
&  \footnotesize {\texttt{DL22}}
&  \footnotesize {\texttt{DL23}} 
& \footnotesize{ \texttt{DL19}}
&  \footnotesize{ \texttt{DL20}}
&  \footnotesize{ \texttt{DL21}}
&  \footnotesize {\texttt{DL22}}
&  \footnotesize {\texttt{DL23}} \\
\midrule
\rowcolor{Gray}  \cellcolor{white} \texttt{DeBERTa-v3}    & 50  & 78.1 & 75.3 & 74.1 & 65.4 & 52.2 & 77.2 & 75.9 & 74.8 & 68.2 & 58.3 & 76.8 & 76.0 & 72.9 & 65.1 & 56.4 \\
&100 &   0.3 & 0.3 & -0.7 & 0.9 & 3.5 & 0.2 & -0.3 & -0.7 & -0.2 & -0.2 & 0.9 & -0.6 & 0.3 & 1.6 & -0.1 \\
&200 & 0.0 & 0.2 & -0.1 & 1.5 & 4.8 & 0.2 & -0.3 & -0.7 & -0.3 & 0.0 & 1.0 & -0.4 & 0.5 & 2.3 & 0.4 \\

\rowcolor{Gray}  \cellcolor{white}  \texttt{ELECTRA} & 50 &  77.5 & 77.1 & 74.5 & 64.1 & 55.5 & 76.9 & 77.7 & 74.7 & 68.1 & 57.9 & 76.7 & 77.2 & 73.3 & 64.5 & 56.9 \\
&100 &  -0.1 & 0.2 & -0.4 & 1.5 & 1.5 & 0.1 & -0.3 & -0.6 & -0.9 & -0.3 & 0.3 & 0.2 & 0.7 & 1.5 & 0.8 \\
&200 & -0.7 & 0.3 & -0.6 & 2.4 & 1.7 & 0.2 & -0.3 & -0.8 & -1.5 & -0.3 & 0.3 & 0.2 & 0.6 & 1.7 & 0.0 \\

\bottomrule
\end{tabular}
}
\label{tab:table_1comptopk}
\end{table*}

\subsection{Cross-Encoders: In-Domain Evaluation}
\label{sec:indomain}
As we provide results for several models, rerankers, and top$_k$ -- $k$ being the number of reranked documents -- we decompose the analysis by first comparing the SPLADE models (Table~\ref{tab:table_1compmodel}), then the rerankers (Table~\ref{tab:table_1compreranker}), and finally the top$_k$ (Table~\ref{tab:table_1comptopk}). In these tables, we fix \textbf{one reference -- in grey -- and display the differences $\Delta$ in performance} between the reference and other results. The full table is also given in Appendix~\ref{appendix:1}.


\subsubsection{Comparing Retrievers} Taking SPLADE-v3-DistilBERT as the reference point, we compare it with SPLADE-v3 and SPLADE-v3-Doc. Considering the baseline row (i.e., first-stage only), we can observe that SPLADE-v3 is usually far more effective --except for the DL19 dataset -- while SPLADE-v3-Doc lags behind. When looking at the number of documents to re-rank, we however see that the gaps between models gradually diminish when increasing $k$.

\subsubsection{Comparing Rerankers} We now compare the DeBERTa-v3 (taken as the reference) and ELECTRA rerankers in Table~\ref{tab:table_1compreranker}. It is difficult to see whether one model is really better than the other: observations vary depending on the dataset. For instance, the model based on ELECTRA performs better for DL20 (and marginally better for DL21) --  but this trend tends to reverse for DL19 or DL22.  

\subsubsection{Impact of $k$} We investigate the impact of $k$ -- the number of documents to re-rank. Using $k=50$ as a reference, we compare it with $k=100$ and $k=200$. We observe that the smallest model (SPLADE-v3-DistilBERT) benefits the most from the increase of $k$, while the others are relatively stable over the $k$ values. This is sensible, as less effective models will tend to retrieve relevant documents at lower ranks. Yet, the impact of $k$ also seems to depend on the dataset used, which makes any general conclusion difficult to make.

To conclude, \emph{there is no general trend to be observed from these in-domain comparisons}. We can see that the best first-stage model \emph{usually} leads to the best end performance, but the rerankers act as expected by narrowing the initial gaps between the three retrievers.

\begin{table*}[ht]
\centering
\caption{
Effectiveness of three SPLADE-v3 models on out-of-domain datasets (BEIR and LoTTE). SPLADE-v3-DistilBERT serves as the reference point (\colorbox{Gray}{grey area}, mean nDCG@10) -- and the comparisons are given in $\Delta$(mean nDCG@10). The baselines correspond to the retrievers only (no re-ranking). 
}
\resizebox{0.8\textwidth}{!}{
\begin{tabular}{ll >{\columncolor{Gray}}c >{\columncolor{Gray}}c >{\columncolor{Gray}}c ccc ccc}
\toprule
  & &\multicolumn{3}{c}{\textbf{SPLADE-v3-DistilBERT}}   & \multicolumn{3}{c}{\textbf{SPLADE-v3}} &  \multicolumn{3}{c}{\textbf{SPLADE-v3-Doc}}   \\ 
\cmidrule(lr){3-5}  \cmidrule(lr){6-8} \cmidrule(lr){9-11} 
\textbf{}
& \textbf{}
& \cellcolor{white}\footnotesize{ \texttt{BEIR}}
& \cellcolor{white} \footnotesize{ \texttt{BEIR-12}}
& \cellcolor{white}  \footnotesize{ \texttt{LoTTE}}
& \footnotesize{ \texttt{BEIR}}
& \footnotesize{ \texttt{BEIR-12}}
&  \footnotesize{ \texttt{LoTTE}}
& \footnotesize{ \texttt{BEIR}}
& \footnotesize{ \texttt{BEIR-12}}
&  \footnotesize{ \texttt{LoTTE}}
\\
\midrule
\texttt{baseline} & & 50.0 & 50.1 & 39.9 & 2.2 & 2.1 & 3.5 & -3.0 & -3.1 & -0.1 \\
\midrule
\textbf{Reranker} & {\textbf{top$_k$}} \\
\texttt{DeBERTa-v3} 
      & 50 & 54.5 & 56.9 & 53.4 & 0.5 & 0.6 & 1.8 & -0.8 & -0.6 & -0.3 \\
&100 & 54.6 & 57.4 & 55.0 & 0.3 & 0.4 & 1.3 & -0.2 & -0.3 & -0.5 \\
&200& 54.9 & 57.5 & 56.1 & 0.1 & 0.4 & 0.7 & -0.3 & 0.0 & -0.4 \\
\texttt{ELECTRA}  
      & 50 & 52.8 & 55.7 & 51.9 & 0.2 & 0.3 & 1.6 & -0.4 & -0.8 & -0.3 \\
&100 & 52.5 & 55.8 & 53.0 & 0.1 & 0.2 & 1.1 & -0.3 & -0.2 & -0.2 \\
&200 & 52.5 & 55.8 & 53.9 & -0.3 & 0.3 & 0.7 & -0.3 & -0.1 & -0.4 \\
\bottomrule
\end{tabular}
}
\label{tab:oodmodels}
\end{table*}

\begin{table*}[ht]
\centering
\caption{
Comparison of DeBERTa-v3 and ELECTRA rerankers for the three SPLADE-v3 models on out-of-domain datasets (BEIR and LoTTE). DeBERTa-v3 serves as the reference point (\colorbox{Gray}{grey area}, mean nDCG@10) -- and the comparisons are given in $\Delta$(mean nDCG@10).
}
\resizebox{0.8\textwidth}{!}{
\begin{tabular}{ll ccc ccc ccc}
\toprule
\multirow{2}{*}{\textbf{Reranker}}& \multirow{2}{*}{\textbf{top$_k$}} &\multicolumn{3}{c}{\textbf{SPLADE-v3-DistilBERT}}   & \multicolumn{3}{c}{\textbf{SPLADE-v3}} &  \multicolumn{3}{c}{\textbf{SPLADE-v3-Doc}}   \\ 
\cmidrule(lr){3-5}  \cmidrule(lr){6-8} \cmidrule(lr){9-11} 
\textbf{}
& \textbf{}
& \footnotesize{ \texttt{BEIR}}
& \footnotesize{ \texttt{BEIR-12}}
&  \footnotesize{ \texttt{LoTTE}}
& \footnotesize{ \texttt{BEIR}}
& \footnotesize{ \texttt{BEIR-12}}
&  \footnotesize{ \texttt{LoTTE}}
& \footnotesize{ \texttt{BEIR}}
& \footnotesize{ \texttt{BEIR-12}}
&  \footnotesize{ \texttt{LoTTE}}
\\

\midrule
\rowcolor{Gray}  \cellcolor{white}\texttt{DeBERTa-v3}    & \cellcolor{white} 50 & 54.5 & 56.9 & 53.4 & 55.0 & 57.5 & 55.2 & 53.8 & 56.3 & 53.1 \\
\rowcolor{Gray}  \cellcolor{white} & \cellcolor{white} 100 & 54.6 & 57.4 & 55.0 & 55.0 & 57.8 & 56.3 & 54.4 & 57.1 & 54.5 \\
\rowcolor{Gray}  \cellcolor{white} & \cellcolor{white} 200 & 54.9 & 57.5 & 56.1 & 54.9 & 57.9 & 56.8 & 54.6 & 57.5 & 55.8 \\
\texttt{ELECTRA} & \\
      & 50 &  -1.7 & -1.2 & -1.5 & -2.0 & -1.5 & -1.7 & -1.3 & -1.4 & -1.5 \\
& 100& -2.1 & -1.6 & -2.0 & -2.3 & -1.7 & -2.2 & -2.1 & -1.4 & -1.7 \\
&200 & -2.4 & -1.7 & -2.2 & -2.7 & -1.7 & -2.2 & -2.4 & -1.8 & -2.2 \\
\bottomrule
\end{tabular}
}
\label{tab:oodreranker}
\end{table*}

\begin{table*}[ht]
\centering
\caption{
Impact of the number of documents to re-rank (top$_k$) on effectiveness on out-of-domain datasets (BEIR and LoTTE). top$_k=50$ serves as the reference point (\colorbox{Gray}{grey area}, mean nDCG@10) -- and the comparisons are given in $\Delta$(mean nDCG@10).
}
\resizebox{0.8\textwidth}{!}{
\begin{tabular}{ll ccc ccc *3c }
\toprule
\multirow{2}{*}{\textbf{Reranker}}& \multirow{2}{*}{\textbf{top$_k$}} &\multicolumn{3}{c}{\textbf{SPLADE-v3-DistilBERT}}   & \multicolumn{3}{c}{\textbf{SPLADE-v3}} &  \multicolumn{3}{c}{\textbf{SPLADE-v3-Doc}}   \\ 
\cmidrule(lr){3-5}  \cmidrule(lr){6-8} \cmidrule(lr){9-11} 
\textbf{}
& \textbf{}
& \footnotesize{ \texttt{BEIR}}
& \footnotesize{ \texttt{BEIR-12}}
&  \footnotesize{ \texttt{LoTTE}}
& \footnotesize{ \texttt{BEIR}}
& \footnotesize{ \texttt{BEIR-12}}
&  \footnotesize{ \texttt{LoTTE}}
& \footnotesize{ \texttt{BEIR}}
& \footnotesize{ \texttt{BEIR-12}}
&  \footnotesize{ \texttt{LoTTE}}
\\
\midrule
\rowcolor{Gray}  \cellcolor{white} \texttt{DeBERTa-v3}    & 50 &  54.48 & 56.95 & 53.36 & 54.97 & 57.52 & 55.20 & 53.72 & 56.33 & 53.09 \\
&100 & 0.14 & 0.43 & 1.65 & 0.00 & 0.26 & 1.09 & 0.66 & 0.72 & 1.44 \\
&200 & 0.40 & 0.57 & 2.78 & -0.03 & 0.39 & 1.64 & 0.89 & 1.17 & 2.68 \\

\rowcolor{Gray}  \cellcolor{white} \texttt{ELECTRA} & 50 & 52.79 & 55.72 & 51.89 & 54.94 & 57.91 & 56.84 & 54.61 & 57.50 & 55.77 \\
&100 & -0.26 & 0.08 & 1.11 & -0.32 & 0.06 & 0.64 & -0.12 & 0.66 & 1.20 \\
&200 & -0.29 & 0.12 & 2.05 & -0.76 & 0.17 & 1.13 & -0.20 & 0.79 & 1.96 \\
\bottomrule
\end{tabular}
}
\label{tab:oodtopk}
\end{table*}

\subsection{Cross-Encoder: Out-of-Domain Evaluation}
\label{sec:ood}

The out-of-domain evaluation brings more contrast to the results. In the following, we provide the same three comparisons as in Section~\ref{sec:indomain}: first-stage models, rerankers models, and the top$_k$. 
The full results are provided in Appendix~\ref{appendix:1}, Table~\ref{tab:splade_rerankerbeirlotte}.

\paragraph{A note on BEIR} Looking at the BEIR dataset, the improvements at first glance were a bit disappointing. With a closer look, we spotted a very strange behavior for the ArguAna dataset. Indeed, the performance tended to (drastically) diminish when increasing the number of documents to re-rank -- e.g., from 50 (nDCG@10) down to approximately 15 for all SPLADE models, leading to almost no improvement for the ELECTRA reranker. This may be explained by the fact that the ArguAna task consists in finding the counter-argument of the ``query''. When discarding  ArguAna (column called BEIR-12), we are ``back on our feet'', and the behavior of BEIR-12 datasets is aligned with LoTTE. We therefore differentiate these two versions of BEIR in our study. 



\subsubsection{Comparing Retrievers} We first compare in Table~\ref{tab:oodmodels}  the impact of the first-stage model. Overall, we can draw similar observations to the in-domain ones: rerankers flatten the differences between models, and with a large enough $k$, most of the differences are below one nDCG@10 point -- while previously reaching up to three points.

\subsubsection{Comparing Rerankers} In Table~\ref{tab:oodreranker}, we further compare rerankers. In out-of-domain, we observe a clear ``winner'': DeBERTa-v3 consistently improves over the ELECTRA-based model.

\subsubsection{Impact of $k$} We additionally show in Table~\ref{tab:oodtopk} that increasing $k$ is a good way to increase the effectiveness of the models. As expected, this is especially true for the weakest models, but even the more effective ones benefit from re-ordering more documents -- especially for the LoTTE dataset.

The findings for the out-of-domain setting are different from the in-domain ones. In this case, a more effective re-ranker (DeBERTa-V3) consistently outperforms its ELECTRA-based counterpart, no matter the retriever. Additionally, increasing the number of documents to re-rank is consistently beneficial (considering the BEIR-12 version).

\begin{table*}[ht]
\centering
\caption{
Evaluation of GPT-based models as zero-shot rerankers on top of SPLADE-v3 (strong baseline) -- nDCG@10. We also report the best TREC runs for each year as comparison points.
}
\resizebox{0.95\textwidth}{!}{
\begin{tabular}{llccccccccccc}
\toprule
& 
& 
& \texttt{DL19}
& \texttt{DL20}
& \texttt{DL21} 
& \texttt{DL22} 
& \texttt{DL23} 
& \texttt{TREC-COVID} 
& \texttt{TREC-NEWS}
& \texttt{Touché-2020}
& \texttt{NovelEval}\\
\midrule
\texttt{SPLADE-v3} & & &  72.3& 75.4  & 70.7 &61.9 &50.6& 74.7& 41.8 &29.3& 70.0 \\
\texttt{Best TREC-DL}  &   &  & 76.4 &  80.3 & 74.9&71.8& 69.9 & - & - & - & -\\
\midrule
\textbf{Reranker} & {\textbf{top$_k$}} & \textbf{W} \\
\texttt{DeBERTa-v3} & 200&  - &  77.6     &   75.6       & 73.4& 67.5& 58.3  &89.2&51.9 & 33.3&82.8\\
\midrule
\textbf{GPT-based} & \\
\texttt{GPT-3.5 Turbo} & 50 &   -  & 73.6  &67.1& 71.0  & 60.7 & 54.1&78.6& 43.4& 24.1&80.2\\
\texttt{GPT-4} & 25  & - &78.5  (0.3) &  77.9 &  75.4& 70.1& 60.2 & 86.9 (0.3)  &49.6& 32.0&90.9\\
\texttt{GPT-4} & 50  &-  & 77.6  & 79.8 &  77.2& 70.0 & 63.0 & 86.9 (0.5)  & 50.2 & 31.6& 87.7\\
\texttt{GPT-4} & 75  & - & 78.8  &78.2    &   77.3 & 70.2 &63.1& 87.5 (0.4)    & - & - & - \\ 
\texttt{GPT-4}  &  50 &   25/10 & -&-&-&-&-&-&81.0&-&30.8&-\\
\texttt{GPT-4}  &  100 &   20/10 & -&-&-&-&-&-&88.2&-&-&-\\ 
\midrule
\textbf{References} & \\
\texttt{RankGPT-4 (BM25)}~\cite{sun_is_2023} &  100& 20/10 &75.59&70.56&-&-&-&-&-&-&-\\
    \texttt{RankGPT-4 (SPLADE++ ED)}~\cite{pradeep_rankzephyr_2023} & 100& 20/10 &74.64&70.76&77.21&71.75&-&87.92  & 53.27&-&90.45\\
   \texttt{RankZephyr (SPLADE++ ED)}~\cite{pradeep_rankzephyr_2023} & 100 &    20/10     &  78.16& 81.59& 75.98& 66.92  &-& 85.35&50.60 &-& 89.34\\
   \texttt{RankZephyr}$\rho$ \texttt{(SPLADE++ ED)}~\cite{pradeep_rankzephyr_2023} & 100 & 20/10        &  78.55& 82.55& 76.88& 66.28  & -&85.66&51.07 &- & -\\
\bottomrule
\end{tabular}
}
\label{tab:ppsdmse}
\end{table*}

\begin{table}[ht]
\centering
\caption{
Comparison between GPT-4 and GPT-4 Turbo (nDCG@10 on DL19, TREC-COVID and TREC-NEWS). Missing points are due to budget limits.   
}
\resizebox{0.5\textwidth}{!}{
\begin{tabular}{llccccc}
\toprule
& 
& 
& \texttt{DL19}
& \texttt{TREC-COVID} 
& \texttt{TREC-NEWS} \\
\midrule
\texttt{SPLADE-v3} & & &  72.3 &  74.7 & 41.8 \\
\midrule
\textbf{Reranker} & {\textbf{top$_k$}}  \\
\texttt{GPT-4}  & 25  &         &78.5 (0.3) & 86.9 (0.3)& 49.6\\
\texttt{GPT-4} & 50   &         & 77.6&86.9 (0.5)& 50.2\\
\texttt{GPT-4} & 75   &         & 78.8&87.5 (0.4)& -\\ 
\midrule
\texttt{GPT-4 Turbo}  &  25     &   & 75.8 & 87.4 &-\\
\texttt{GPT-4 Turbo}  &  50     & & 77.5 (0.4) &-&48.3  \\
\texttt{GPT-4 Turbo}  &  100    & & 76.8 &85.2& -\\
\bottomrule
\end{tabular}
}
\label{tab:gpt4turbo}
\end{table}

\subsection{LLM as Rerankers}\label{sec:4-3}

We now discuss the evaluation of LLMs as rerankers. First, we focus on OpenAI models used by {\it Sun et al.}~\cite{sun_is_2023}: GPT-3.5 Turbo and GPT-4. 
Unfortunately, the cost of using such models prevents us from conducting extensive experiments. For some configurations, we have conducted 3 runs to asses the LLMs variations to sampling, the standard deviation being around 0.2/0.5 point for the nDCG@10 measure. These results are indicated in Table~\ref{tab:ppsdmse} with the format: AVG (STD). 

We present in Table~\ref{tab:ppsdmse} the evaluation of GPT-3.5 Turbo and GPT-4 re-ranking documents from SPLADE-v3, for the in-domain datasets and for the following out-of-domain datasets: TREC-COVID, TREC-NEWS, Touché-2020 and NovelEval. As a comparison, we also report the DeBERTa-v3 results from Table~\ref{tab:table_1compreranker}. 

As we can see, GPT-3.5 Turbo is able to improve performance when reranking, but falls short compared to DeBERTa-v3. In some cases, it even degrades the results of the retriever (DL20, DL22, Touché). It is interesting to note that \textbf{this degradation is not visible when BM25 is used as a first-stage} \cite{sun_is_2023} (see DL20 results).  
On the other hand, GPT-4 performs well -- on par with DeBERTa-v3 -- and even better for some datasets (especially DL23 and NovelEval).

We tested several configurations with GPT-4: various top$_k$ documents ($k \in \{25, 50, 75, 100\}$), and with or without the sliding window mechanism.  
These various configurations used for GPT-4 tend to show that the sliding window mechanism does not seem necessary: the shortening mechanism (Section \ref{sec:shorten}) often provides results on par with (or even better than) the sliding window mechanism. This observation may be different with BM25, where more documents ($k$=100) may be needed to get similar results. Therefore, a more effective retriever makes the job of rerankers easier. 

To extend the comparison and add more reference points, we added results from some TREC participants: these results are usually very competitive, but obtained by combining a set of various models. We also add the recent results obtained by RankZephyr \cite{tang_found_2023}, which \emph{fine-tunes} a Zephyr model for the listwise reranking task. The $\rho$ version corresponds to the RankZephyr model progressively reranking the input three times using a SPLADE++ ED model \cite{formal2022distillation}. This model in general performs very well, especially for the DL20 dataset, but is on par with DeBERTa-v3 reranker for many other datasets. Please note that, for these results, the retriever is slightly less effective than SPLADE-v3, so results are not entirely comparable -- but are given as another comparison point (The RankZephyr model is not publicly available as of January, 14 2024).

We also test the recent GPT-4 Turbo in  Table~\ref{tab:gpt4turbo}: it has a longer prompt length (100$k$ instead of 8$k$), but the results are a bit underwhelming. Especially, the performance does not increase with the number of documents to re-rank: it performs better with $k$=25 documents than with $k$=100 documents. Therefore, even if a model can deal with longer prompts, the ``naive'' strategy of increasing $k$ might be suboptimal.

\subsection{Detailed Comparison on TREC-COVID}

We now focus on the TREC-COVID dataset \cite{10.1145/3451964.3451965}, and further compare our DeBERTa-v3 reranker with the OpenAI LLMs, as well as the ``open'' ones. The results are presented in Table \ref{tab:covid}.

\subsubsection{Comparing DeBERTa-v3 and OpenAI LLMs}
\begin{table}[ht]
\centering
\caption{
Comparison of cross-encoders, open and closed LLMs on TREC-COVID (nDCG@10). $|d|$ indicates the truncation length, and ``title'' indicates whether this field is used at evaluation (\faCheck
) or not (\faTimes).
}
\resizebox{0.5\textwidth}{!}{
\begin{tabular}{lccccc}
\toprule
 
& 
& 
& 
& \texttt{title} 
&   \texttt{nDCG@10} \\
\midrule
\texttt{SPLADE-v3} &     &  &  &  \faCheck & 74.7\\
\midrule
\textbf{Reranker} & {\textbf{top$_k$}} & $|d|$ & \textbf{W} \\
\rowcolor{Gray} \texttt{DeBERTa-v3} & 200 &512 &- & \faCheck & 89.2\\
\texttt{DeBERTa-v3} & 50 & 64&- & \faCheck &82.6 \\
\texttt{DeBERTa-v3} & 100 &64 &- & \faCheck  &80.9\\
\texttt{DeBERTa-v3} & 200 & 64&- & \faCheck &79.1\\
\texttt{DeBERTa-v3} & 50 & 512&- & \faTimes & 84.5\\
\texttt{DeBERTa-v3} & 100 & 512&- & \faTimes &84.1\\
\texttt{DeBERTa-v3} & 200 & 512&- & \faTimes &84.6\\
\texttt{DeBERTa-v3} & 50 & 64&- & \faTimes &71.2 \\
\texttt{DeBERTa-v3} & 100 &64 &- &\faTimes  &67.8\\
\texttt{DeBERTa-v3} & 200 & 64&- & \faTimes &65.5\\
\midrule
\textbf{GPT-based} & \\
\texttt{GPT-3.5 Turbo}  & 50   &512  &   - & \faCheck &78.6\\
\texttt{GPT-4}          & 25 &64  & -  & \faCheck & 86.2\\
\texttt{GPT-4}          & 25 & 64 & -  & \faTimes &82.3 \\
\texttt{GPT-4}          & 25 & 79 & -  & \faTimes &86.2 \\
\texttt{GPT-4}          & 25 & 79 & -  & \faCheck &86.9 (0.2) \\
\texttt{GPT-4}          & 50 &  64& -  & \faCheck & 86.9 (0.5)\\
\texttt{GPT-4}          & 75 & 79&  - & \faCheck & 87.5 (0.4)\\
\texttt{GPT-4}          &100 & 79 & 20/10 &\faCheck& 88.2 \\
\texttt{GPT-4 Turbo}     & 25  & 214  &  - & \faCheck & 87.3 \\
\texttt{GPT-4 Turbo}     & 100  & 214 & -  & \faCheck & 85.2 \\
\midrule
\textbf{Other LLMs} & \\
\texttt{SOLAR 10.7B}       & 15  &  64 &  - & \faCheck &76.4\\
\texttt{SOLAR 10.7B}       & 50  & 64  & 10/5  & \faCheck &77.6\\
\texttt{Yi-34B-Chat}          & 15  &  64 &  - & \faCheck &77.8\\
\texttt{Yi-34B-Chat}          & 25  & 64  & 10/5  & \faCheck &81.1\\
\texttt{Yi-34B-Chat}          &50   & 64   & 10/5  &\faCheck& 82.3 \\
\texttt{Llama-70B-chat} & 15& 64  &- & \faCheck&77.8\\
\texttt{Llama-70B-chat} & 25& 64  &- & \faCheck&76.8\\
\midrule 
\textbf{References} & \\
\texttt{RankZephyr} \cite{pradeep_rankzephyr_2023} & 100 & &20/10& \faCheck &   85.3\\
\texttt{RankZephyr} $\rho$  \cite{pradeep_rankzephyr_2023} &100&&20/10& \faCheck & 85.7   \\

\end{tabular}
}
\label{tab:covid}
\end{table}

For DeBERTa-v3, we consider different cut-offs $k$ as well as document truncation lengths. For GPT-4, we additionally tested the sliding window mechanism. Overall, the cross-encoder DeBERTa-v3 obtains the best results, by reranking $k$=200 documents.

Nevertheless, we observe one intriguing result with GPT-4: it is able to perform extremely well when considering very short documents (to fit the prompt length -- see Section~\ref{sec:shorten}). Here, documents are truncated from an average of 214 tokens down to 79 tokens approximately. Yet, the results are very competitive (around 87 nDCG@10). Applying the same strategy for the DeBERTa-v3 reranker has much more impact on effectiveness (from approximately 89 nDCG@10 down to 80). 

We further investigate the impact of the \texttt{title} on different re-rankers: this information concatenated at the beginning of the document content helps the LLMs, especially if the documents are shorter. We also note that, similarly to the document length, DeBERTa-v3 is more sensitive than GPT-4  regarding the absence/presence of a title: it loses $\downarrow$ 5 nDCG@10  points, while GPT-4's effectiveness only drops by $\downarrow$ 1 nDCG@10 point. The effect is especially visible when the documents are shorter. Note that it is difficult to generalize findings about the impact of titles, as they may be specific to the TREC-COVID collection. 

Regarding the reranking depth, we have seen that increasing $k$ is usually beneficial for DeBERTa-v3 in the out-of-domain setting (Section~\ref{sec:ood}). On TREC-COVID, it goes up to 89 nDCG@10. GPT-4 achieves very close results (87 nDCG@10),  with a smaller $k$ -- i.e., $k=25$ or $k=50$ -- but does not seem to be able to leverage higher re-ranking depths -- even with the sliding window strategy. The sliding mechanism marginally improves the result (88.2  nDCG@10) while requiring twice as much time.

\begin{table}[ht]
    \centering
        \caption{Cascading pipelines with LLMs (nDCG@10 on TREC DL23).}
    \begin{tabular}{lc}
    \toprule
    \textbf{Ranking pipeline} &\texttt{nDCG@10} \\
    \midrule
     \rowcolor{Gray} \texttt{SPLADE-v3} $\gg$ \texttt{DeBERTa-v3} ($k=200$)    & 58.3 \\
     \texttt{SPLADE-v3} $\gg$ \texttt{GPT-4} ($k=25$)   & 60.2 \\
    \texttt{SPLADE-v3} $\gg$ \texttt{GPT-4} ($k=50$)  & 63.0 \\
     \texttt{SPLADE-v3} $\gg$ GPT-4 ($k=75$)    &   63.1 \\
      \texttt{SPLADE-v3}  $\gg$ \texttt{DeBERTa-v3} ($k=200$) $\gg$ \texttt{GPT-4} ($k=25$)   &63.7  \\
     \bottomrule
    \end{tabular}
    \label{tab:pipeline}
\end{table}

\subsubsection{Comparing Closed and Open Models}

We also report in Table~\ref{tab:covid} results from open LLMs: SOLAR 10.7$B$, Yi-34$B$-Chat \cite{ai2024yi}, and Llama-70$B$-chat \cite{touvron2023llama}, used as zero-shot re-rankers. The results are similar to those reported by {\it Qin et al.}~\cite{qin_large_2023}: very underwhelming -- the generated outputs being too far away from the expected format. However, we found that the amount of text  fed into the prompt has to be reduced (through $k$) compared to the OpenAI models to achieve interesting results. Indeed, some models (e.g., Yi-34$B$-Chat) can achieve decent results (up to $82$ nDCG@10), especially compared to GPT-3.5 Turbo, but $k$ must be smaller (around 10-15).  Interestingly the sliding window mechanism improves the results, allowing the use of a greater number of documents (from 15 to 50).

Since we know the training material used by these open models, especially for the instruct versions (using RLHF or DPO), we can reasonably infer that the ranking capabilities of the closed models also appear in the open ones, and that this capability is not due to some specific task-oriented data. 

\subsection{Reranking Pipeline}

GPT-4 can be used to iteratively rerank a set of documents \cite{sun_is_2023,pradeep_rankzephyr_2023}. Due to budget limits, we do not report such experiments here, but rather assess whether LLMs-based rerankers could be used to re-order results from cross-encoders in a cascade scenario (see the winner systems of the TREC 2023 DL track \cite{craswell2024overview}). Table~\ref{tab:pipeline} reports the gain obtained by reranking the DeBERTa-v3 output: with only 25 documents, it outperforms GPT-4 results obtained with more documents ($k$=50 or $k$=75). This indicates that effective re-rankers can also be used to select candidates to be re-ranked by costly LLMs in a cascading IR pipeline. LLM-based re-rankers are therefore not necessarily bound to replace cross-encoders, but rather act as new ``contenders'' in the effectiveness-efficiency spectrum of IR systems.  

\section{Concluding Remarks}

We have evaluated several reranking methods for strong SPLADE retrievers. If the effectiveness hierarchy between the first-stage models is preserved, the use of cross-encoder-based rerankers has an interesting smoothing effect, especially when increasing the number of reranked documents. 
In the in-domain setting, it is difficult to see some strong differences between evaluated rerankers. However, in out-of-domain, the gap between approaches widens. Overall, increasing the number of documents to re-rank has a positive impact on the final effectiveness. We additionally observe that effective cross-encoders (coupled with strong retrievers) are able to outperform all LLMs -- except GPT-4 on some datasets -- while being far more efficient.

We also evaluated (zero-shot) LLM-based rerankers using OpenAI and open models. While the open LLMs are on par with GPT-3.5 Turbo, GPT-4 exhibits a very surprising ability to re-rank documents. Despite their effectiveness, both the inefficiency and large cost of the models prevent them from being used in retrieval systems. They however constitute a new contender in the effectiveness-efficiency spectrum.  





\bibliographystyle{ACM-Reference-Format}
\balance{}
\bibliography{ref}

\newpage
\onecolumn
\section*{Appendix}
\label{appendix:1}

In this Appendix, we provide the original full Tables from which we derived the Tables used in Sections~\ref{sec:indomain}  and~\ref{sec:ood} for comparing various viewpoints. We also provide some examples of prompts and outputs used with the LLMs. 

\subsection*{Prompts}

We use the prompt shown in Table~\ref{tab:prompt}, inspired by {\it Sun et al.}~\cite{sun_is_2023} for all experiments done with LLMs. 
Note that the post-processing of the incorrect output is able to extract a well-formed output.

\begin{table}[h]
 \centering
        \caption{Prompt template used.}
        \resizebox{1\textwidth}{!}{
\begin{tabular}{l|l}
\toprule
\texttt{System} & You are an intelligent assistant that can rank passages based \\
       &      on their relevancy to the query. Only response the ranking results, do not say any word \\
       &      nor explain. \\
\texttt{User} & I will provide you with \{num\} passages, each indicated by number identifier []. \\
       &  Rank the passages based on their relevance to query: \{query\}. \\
\texttt{Assistant} & Okay, please provide the passages.\\
\texttt{User} & [1] Impact of population mask wearing on Covid-19 post lockdown COVID-19, caused by SARS-CoV2 [...] \\
\texttt{Assistant} & Received passage [1].\\
 & (other documents) \\
\texttt{User} & Search Query: \{query\}. Rank the \{num\} passages above based on their relevance \\
   & to the search query. The passages should be listed in descending order using identifiers. \\
  &  The most relevant passages should be listed first. The output format should be [] > [], e.g., [2] > [1]. \\
  &  Only response the ranking results, do not say any word nor explain and write only the sorted list. \\
  &  Rank now!\\     
\midrule      
Example of correct output& [4] > [1] > [5] > [6] > [9] > [8] > [3] > [10] > [2] > [7] \\
Example of incorrect output&  [1] A critical evaluation of glucocorticoids in the treatment of severe COVID-19\\
&[4] A critical evaluation of glucocorticoids in the management of severe COVID-19\\
&[3] Dexamethasone for COVID-19? Not so fast\\
&[8] Rapid Radiological Worsening and Cytokine Storm Syndrome in COVID-19 Pneumonia\\
&[9] Reducing dexamethasone antiemetic prophylaxis during the COVID-19 pandemic: recommendations from \\
& Ontario, Canada\\
&[6] Multiple Myeloma in the Time of COVID-19\\
&[5] Transcatheter drug delivery through bronchial artery for COVID-19: is it fiction or could it come true?\\
&[7] Multiple Myeloma in the Time of COVID-19\\
&[10] Reducing de\\
\end{tabular}}
\label{tab:prompt}
\end{table}

\begin{table}[ht]
    \centering
        \caption{HuggingFace model names}
    \begin{tabular}{l|l}
    \toprule
       \textbf{Model}  &  \textbf{HuggingFace model name}\\
       \midrule
       SPLADE-v3 & naver/splade-v3 \\
       SPLADE-v3-DistilBERT &naver/splade-v3-distilbert \\ 
       SPLADE-v3-Doc & naver/splade-v3-doc\\
       DebertaV3  & naver/trecdl22-crossencoder-debertav3\\ 
       Electra  & naver/trecdl22-crossencoder-electra \\     
       Yi   &01-ai/Yi-34B-Chat  \\
       SOLAR &Upstage/SOLAR-10.7B-Instruct-v1.0 \\
        & \\
       \bottomrule
    \end{tabular}
    \label{tab:models}
\end{table}

\begin{table*}[ht]
\centering
\caption{
In domain Evaluation (nDCG@10) of various rerankers with various SPLADE models 
}
\resizebox{0.9\textwidth}{!}{
\begin{tabular}{l|l ||ccccc|| ccccc|| ccccc|| ccccc}
\toprule
\multirow{2}{*}{\textbf{Reranker}}& \multirow{2}{*}{\textbf{top$_k$}} &\multicolumn{5}{c}{\textbf{SPLADE-v3}}   & \multicolumn{5}{c}{\textbf{SPLADE-v3-Doc}} &  \multicolumn{5}{c}{\textbf{SPLADE-v3-DistilBERT}}   \\ 
\cmidrule(lr){3-7}  \cmidrule(lr){8-12} \cmidrule(lr){13-17} 
\textbf{}
& \textbf{}
& \footnotesize{ \textbf{DL19}}
&  \footnotesize{ \textbf{DL20}}
&  \footnotesize{ \textbf{DL21}}
&  \footnotesize {\textbf{DL22}}
&  \footnotesize {\textbf{DL23}} 
& \footnotesize{ \textbf{DL19}}
&  \footnotesize{ \textbf{DL20}}
&  \footnotesize{ \textbf{DL21}}
&  \footnotesize {\textbf{DL22}}
&  \footnotesize {\textbf{DL23}} 
& \footnotesize{ \textbf{DL19}}
&  \footnotesize{ \textbf{DL20}}
&  \footnotesize{ \textbf{DL21}}
&  \footnotesize {\textbf{DL22}}
&  \footnotesize {\textbf{DL23}} \\
\midrule
SPLADE-v3 & $-$ & 72.26 & 75.36&70.73& 61.93&50.75 & 71.53&70.31&64.99&51.54&46.42&75.24&74.42&65.25&55.08&46.65\\
\midrule
DebertaV3    & 50 &77.24&75.88&74.71&\textbf{68.16}&\textbf{58.32}&76.85&75.96&72.89&65.16&56.39&78.15&75.26&74.06&65.44&52.24\\
      & 100&   77.43&75.55&73.96&67.97&58.09&\textbf{77.73}&75.38&73.24&66.76&56.32 & 78.49&75.53&73.37&66.31&55.76\\
     & 200& 77.47     & 75.56    &73.96 &67.82 &58.33 &77.82 &75.52&73.37&67.47&56.82&78.17& 75.50& 73.96& 66.98&57.03\\
\midrule
Electra & 50 &76.92&\textbf{77.74}&\textbf{74.75}&68.08&57.93&76.68&77.22&73.27&64.49&56.95&77.50&77.15&74.55&64.13&55.51\\
      & 100&77.02&77.45&74.11&67.22&57.60&77.02&77.38&74.00&66.03&57.74&77.44&77.35& 74.18&65.59&56.98 \\
      & 200 & 77.17   &  77.46  &73.91 & 66.15  & 56.97&76.99 &77.46 &73.91 &66.15 &56.97 &76.76 &77.47&73.99& 66.58&57.25\\
\bottomrule
\end{tabular}
}
\label{tab:splade_rerankertrec}
\end{table*}

\begin{table*}[ht]
\centering
\caption{
Out-of-Domain Evaluation (nDCG@10). BEIR-12 corresponds to the BEIR-13 dataset discarding Arguana.
}
\resizebox{0.7\textwidth}{!}{
\begin{tabular}{ll ccc ccc ccc}
\toprule
\multirow{2}{*}{\textbf{Reranker}}& \multirow{2}{*}{\textbf{top$_k$}} &\multicolumn{3}{c}{\textbf{SPLADE-v3}}   & \multicolumn{3}{c}{\textbf{SPLADE-v3-Doc}} &  \multicolumn{3}{c}{\textbf{SPLADE-v3-DistilBERT}}   \\ 
\cmidrule(lr){3-5}  \cmidrule(lr){6-8} \cmidrule(lr){9-11} 
\textbf{}
& \textbf{}
& \footnotesize{ \textbf{BEIR}}
& \footnotesize{ \textbf{BEIR-12}}
&  \footnotesize{ \textbf{LOTTE}}
& \footnotesize{ \textbf{BEIR}}
& \footnotesize{ \textbf{BEIR-12}}
&  \footnotesize{ \textbf{LOTTE}}
& \footnotesize{ \textbf{BEIR}}
& \footnotesize{ \textbf{BEIR-12}}
&  \footnotesize{ \textbf{LOTTE}}
\\
\midrule
baseline & -&52.19&52.20& 43.33 & 47.0&46.98&39.75&50.0&50.12&39.87\\
\midrule
\multicolumn{11}{l}{\textbf{debertav3}}\\
      & 50 & 54.97&57.52&55.20&53.72&56.33&53.09&54.48&56.95&53.36 \\
      & 100& 54.97&57.78& 56.29&54.38 &57.05&54.53&54.62&57.38&55.01\\
      & 200& 54.94 &57.91& 56.84&54.61&57.50  &55.77&54.88&57.52& 56.14   \\
Electra & \\
      & 50 &52.99&55.99&53.48  &52.38 &54.95&51.62   &52.79  &55.72&51.89\\
      & 100&52.67 &56.05&54.12& 52.26  &55.61&52.82   & 52.53 &55.80 &53.00\\
      & 200& 52.23 &56.16 &54.61    &52.18 & 55.74  &53.58& 52.50&55.84& 53.94\\
\bottomrule
\end{tabular}
}
\label{tab:splade_rerankerbeirlotte}
\end{table*}

\end{document}